\documentstyle[12pt]{article}
\newcommand{\beq}{\begin{equation}}             
\newcommand{\eeq}{\end{equation}}               
\newcommand{\bqry}{\begin{eqnarray}}            
\newcommand{\eqry}{\end{eqnarray}}              
\newcommand{\bqryn}{\begin{eqnarray*}}          
\newcommand{\eqryn}{\end{eqnarray*}}            
\newcommand{\preprint}[1]{\begin{table}[t]      
            \begin{flushright}                  
            \begin{large}{#1}\end{large}        
            \end{flushright}                    
            \end{table}}                        
\newcommand{\PD}[2]                             
    {\frac{\partial^{#2}}{\partial #1^{#2}}}    
\begin{document}
\preprint{LA-UR-98-1915}
\title{Hadron Spectroscopy in Regge Phenomenology\thanks{Presented at the 
First International Conference on Parametrized Relativistic Quantum Theory, 
PRQT '98, Houston, Texas, USA, Feb 9-11, 1998}}
\author{\\ L. Burakovsky\thanks{E-mail: BURAKOV@QMC.LANL.GOV} \
\\  \\  Theoretical Division, T-8 \\  Los Alamos National Laboratory \\ Los
Alamos NM 87545, USA}
\date{ }
\maketitle
\begin{abstract}
We show that linear Regge trajectories for mesons and baryons, and the cubic
mass spectrum associated with them, determine expressions for the hadron masses
in terms of the universal Regge slope $\alpha ^{'}$ alone. The ground state 
hadron masses as calculated from these expressions are in excellent agreement 
with experiment for $\alpha ^{'}=0.85$ GeV$^{-2}.$ 
\end{abstract}

\bigskip
\section{Introduction}
It is well known that the hadrons composed of light $(n(=u,d),s)$ quarks 
populate linear Regge trajectories; i.e., the square of the mass of a state 
with orbital momentum $\ell $ is proportional to $\ell:$ $M^2(\ell )=\ell /
\alpha ^{'}\;+$ const, where the slope $\alpha ^{'}$ only very weekly depends 
on the flavor content of the states lying on the corresponding trajectory: 
$\alpha ^{'}_{n\bar{n}}=0.88$ GeV$^{-2},$ $\alpha ^{'}_{s\bar{n}}=0.85$ GeV$^{
-2},$ $\alpha ^{'}_{s\bar{s}}=0.81$ GeV$^{-2};$ it therefore may be taken as a
universal slope in the light quark sector, $\alpha ^{'}\approx 0.85$ GeV$^{
-2}.$ In this respect, the hadron masses as populating collinear trajectories 
exhibit a universal behavior governed by the only parameter $\alpha ^{'}.$ It 
is therefore very interesting to ask whether the hadron masses can be expressed
in terms of this parameter alone. The answer to this non-trivial at first 
glance question turns out positive, if one resorts to the notion of the hadron
mass spectrum. 

The idea of the spectral description of a strongly interacting gas which is a 
model for hot hadronic matter was suggested by Belenky and Landau \cite{BL} 
and consists in considering the unstable particles (resonances) on an equal 
footing with the stable ones in the thermodynamic quantities, by means of the 
resonance spectrum; e.g., the expression for pressure in such a resonance gas 
reads (in the Maxwell-Boltzmann approximation)
\beq
p=\sum _ig_i\;p(m_i)=\int _{M_l}^{M_h}dm\;\tau (m)\;p(m),\;\;\;p(m)=\frac{T^2
m^2}{2\pi ^2}K_2\left( \frac{m}{T}\right) ,
\eeq
where $M_l$ and $M_h$ are the masses of the lightest and heaviest species,
respectively, and $g_i$ are particle degeneracies. 

Phenomenological studies \cite{phen,linNPA} have suggested that the cubic 
density of states, $\tau (m)\sim m^3,$ for each isospin and hypercharge 
provides a good fit to the observed hadron spectrum. Let us demonstrate here 
that this cubic spectrum is intrinsically related to collinear Regge 
trajectories (for each isospin and hypercharge).

\subsection{Mass spectrum of linear Regge trajectories}
It is very easy to show that the mass spectrum of an individual Regge 
trajectory is cubic. Indeed, consider, e.g., a model linear trajectory with 
negative intercept: 
\beq
\alpha (t)=\alpha ^{'}\;\!t-1.
\eeq
The integer values of $\alpha (t)$ correspond to the states with integer spin, 
$J=\alpha (t_J),$ the masses squared of which are $m^2(J)=t_J.$ Since a 
spin-$J$ state has multiplicity $2J+1,$ the number of states with spin 
$0\leq J\leq {\cal J}$ is
\beq
N({\cal J})=\sum _{J=0}^{\cal J}(2J+1)=({\cal J}+1)^2=\alpha ^{'2}m^4(
{\cal J}),
\eeq
in view of (2), and therefore the density of states per unit mass interval   
(the mass spectrum) is
\beq
\tau (m)=\frac{dN(m)}{dm}=4\alpha ^{'2}m^3.
\eeq
It is also clear that for a finite number of collinear trajectories, the 
resulting mass spectrum is
\beq
\tau (m)=4N\alpha ^{'2}m^3,
\eeq
where $N$ is the number of trajectories, and does not depend on the numerical 
values of trajectory intercepts, as far as its asymptotic form $m\rightarrow
\infty $ is concerned.

\subsection{Mass spectrum of an individual hadronic multiplet}
Similar to the cubic spectrum of the family of Regge trajectories, 
phenomenological studies have suggested that the mass spectrum of an individual
hadronic multiplet is linear, for both mesons \cite{linmes} and baryons 
\cite{linbar}. It turns out that the form (5) of the cubic spectrum of the 
family of collinear Regge trajectories indeed leads to the linear mass spectrum
of an individual multiplet, and allows one to establish the normalization 
constant of this linear spectrum, as follows:

Consider the family of hadronic multiplets with spin 0,1,..., which populate
collinear trajectories. Then the total number of states can be obtained in two
ways: summing up individual trajectories for every fixed value of isospin, or
summing up individual multiplets for every fixed value of spin. Either way 
should lead to the cubic spectrum, as discussed above. In the case of meson
multiplets (similar analysis may be done in the case of baryon multiplets, of
course), in Eq. (1) $g_i=(2J_i+1)(2I_i+1)^{'},$ where $J_i$ and $I_i$ are 
the values of individual spin and isospin, respectively (``$^{'}$'' means that
for $I_i=1/2,$ the above expression for $g_i$ should be multiplied by 2). 
Then, in view of (1),(5),
\beq
p=\sum _i(2J_i+1)(2I_i+1)^{'}p(m_i)\simeq 4N\alpha ^{'2}\int dm\;m^3\;p(m).
\eeq
Since also $J_i\simeq \alpha ^{'}m_i^2,$ it follows from the above expression
that\footnote{This results may be rigorously proven by the use of, e.g., the
Euler-Maclaurin summation formula which relates a sum to an integral.}
\beq
\sum _i(2I_i+1)p(m_i)\simeq 2N\alpha ^{'}\int dm\;m\;p(m),
\eeq
and since Eq. (7) corresponds to an individual meson multiplet (with fixed 
spin $J_i),$ one sees that the mass spectrum of an individual meson multiplet 
is indeed linear, and its normalization constant is $C=2N\alpha ^{'}.$ 

Now we are ready to show how the cubic spectrum of the family of multiplets and
the linear spectrum of an individual multiplet can predict the masses of the
states. 

\section{Particle spectroscopy}
Let us start with meson spectroscopy. To establish the masses of the states in 
the model of collinear Regge trajectories discussed above, one has to know the
intramultiplet mass splitting $m^2_{I=1/2}-m^2_{I=1}$ and the mass of the
lowest-lying isovector, $m_{I=1}.$ The former can be easily found with the 
help of (7), for 9 isospin degrees of freedom of a meson nonet placed in 
the mass interval\footnote{We assume that the remaining ninth isoscalar 
belongs to this interval; As established in \cite{linmes}, for idealized meson
nonets, its mass is equal to $(2m^2_{I=1/2}+m^2_{I=1})/3$ which lies between
$m_{I=1}^2$ and $m_{I=1/2}^2<m^2_{I=0}.$} $(m_{I=1},\;\!m_{I=0}),$ 
with\footnote{This follows from the standard Gell-Mann--Okubo mass formula
$3m^2_{I=0}+m^2_{I=1}=4m^2_{I=1/2}$ which is also provided by the linear 
mass spectrum of a meson octet \cite{linmes}.} 
$m^2_{I=0}-m^2_{I=1}=4/3\;(m^2_{I=1/2}-m^2_{I=1})\equiv 4/3\;\triangle :$
\beq
9=2N\alpha ^{'}\int _{m_{I=1}}^{m_{I=0}}dm\;m=2N\alpha ^{'}\;\!\frac{1}{2}\;\!
\frac{4}{3}\triangle ;
\eeq
therefore
\beq
\triangle =\frac{27}{4N\alpha ^{'}}.
\eeq
To determine the number of collinear trajectories, we note that there are
four different meson multiplets for every partial wave, except for $S$-wave, 
which in the standard spectroscopic notation are\footnote{In a constituent 
quark model, these multiplets correspond to spin-singlet and spin-triplet 
states of a bound system of two quarks.}
\bqryn
^1S_0 & ^1P_1\;\;\; ^1D_2\;\;\; ^1F_3\;\;\ldots  \\ 
      & ^3P_0\;\;\; ^3D_1\;\;\; ^3F_2\;\;\ldots  \\
      & ^3P_1\;\;\; ^3D_2\;\;\; ^3F_3\;\;\ldots  \\
^3S_1 & ^3P_2\;\;\; ^3D_3\;\;\; ^3F_4\;\;\ldots  
\eqryn
(note that two missing $S$-wave nonets can be replaced by the radial 
excitations of $^1S_0$ and $^3S_1),$ each of which contains 9 isospin states;
therefore, the total number of different collinear meson trajectories is 
$$N=4\times 9=36.$$ Hence, as follows from (9),
\beq
\triangle =\frac{3}{16\alpha ^{'}}.
\eeq
It is well known that two isoscalar states of an idealized bare meson nonet 
mix with each other to form the physical states the masses of which are 
\cite{linmes}
\beq
m^2_{I=0'}=m^2_{I=1},\;\;\;m^2_{I=0''}=2m^2_{I=1/2}-m^2_{I=1}. 
\eeq
Therefore, one has 
\beq
m^2_{K^\ast }=m^2_\rho +\frac{3}{16\alpha ^{'}},\;\;\;m^2_{\phi }=m^2_\rho +
\frac{3}{8\alpha ^{'}},\;\;\;{\rm etc.,}
\eeq
and also
\beq
m^2_K=m^2_\pi +\frac{3}{16\alpha ^{'}}.
\eeq

It is widely believed that pseudoscalar mesons are the Goldstone bosons of
broken SU(3)$\times $SU(3) chiral symmetry of QCD, and that they should be 
massless in the chirally-symmetric phase. Therefore, it is not clear how well 
would the framework that we discuss here be suitable for the description of
the pseudoscalar nonet. Indeed, as we have tested in \cite{BHNPA}, this nonet
is {\it not} described by the linear spectrum. Moreover, pseudoscalar mesons 
are extremely narrow (zero width) states to fit into a resonance description. 
Probably, the manifestly covariant framework cannot predict the mass of the 
pion, although the formula (13) is consistent with data, as we shall see below.
 
Thus, the resonance description should start with vector mesons, and the cubic
spectrum of a linear trajectory enables one to determine the mass of the 
$\rho $ meson, as follows:
 
Since the $\rho $ meson has the lowest mass which the resonance description 
starts with, let us locate this state by normalizing the $\rho $ trajectory
to one state in the characteristic mass interval
$(\sqrt{m^2_\rho -1/(2\alpha ^{'})},\;\!\sqrt{m^2_\rho +1/(2\alpha ^{'})}).$ 
With the cubic spectrum (4) of a linear trajectory, one has\footnote{Since 
the $\rho $ trajectory starts with a spin-1 isospin-1 state $(\rho ),$ it 
corresponds to the spectrum $\tau (m)=9\times 4\alpha ^{'2}m^3.$ There is 
therefore no difference in normalizing this trajectory to 9 states, or (4) 
to one state, in the vicinity of the $\rho $ mass.}
\beq
1=4\alpha ^{'2}\int _{\sqrt{m^2_\rho -1/(2\alpha ^{'})}}^{\sqrt{m^2_\rho +
1/(2\alpha ^{'})}}m^3\;dm=2\alpha ^{'}m^2_\rho ;
\eeq
therefore 
\beq
m^2_{\rho }=\frac{1}{2\alpha ^{'}},
\eeq
and, through (12),
\beq
m^2_{K^\ast }=\frac{11}{16\alpha ^{'}},\;\;\;m^2_{\phi }=\frac{7}{8\alpha ^{'}}
,\;\;\;{\rm etc.}
\eeq

Similar analysis can be easily done for baryons. Here, for brevity, we skip 
this analysis and only refer to \cite{linbar} where preliminary discussion on 
the baryon spectroscopy can be found. Let us just write down the final 
expressions:
\beq
m^2_N=\frac{3}{4\alpha ^{'}},\;\;\;m^2_{\Sigma ^{'}}=\frac{9}{8\alpha ^{'}},
\;\;\;m^2_\Xi =\frac{3}{2\alpha ^{'}},
\eeq
\beq
m^2_\Delta =\frac{5}{4\alpha ^{'}},\;\;\;m^2_{\Sigma ^\ast }=\frac{13}{8\alpha
^{'}},\;\;\;m^2_{\Xi ^\ast }=\frac{2}{\alpha ^{'}},\;\;\;m^2_\Omega =\frac{
19}{8\alpha ^{'}},\;\;\;{\rm etc.}
\eeq
In (17), $m^2_{\Sigma ^{'}}\equiv (m^2_\Lambda +m^2_\Sigma )/2$ \cite{linbar}.
  
It is seen in (17),(18) that the mass squared splitting within an 
individual baryon multiplet is twice as large as that for an individual meson 
multiplet; e.g., $m^2_{\Sigma ^\ast }-m^2_\Delta =3/(8 \alpha ^{'}),$ as 
compared to (12),(13). The mass squared splitting between multiplets which
differ by one unit of spin remains, however, the same: since $m_\pi \ll m_\rho
,$ it follows from (15),(17),(18) that $m^2_\rho -m^2_\pi \approx m^2_\rho
=1/(2\alpha ^{'})=m^2_\Delta -m^2_N.$ Also, the relation $m_N^2=3/2\;m_\rho ^
2,$ as follows from (15),(17), is definitely related to the valence quark 
structure interpretation of the two states.

\subsection{Comparison with data}
Now we wish to compare the formulas (13),(15)-(18) with available 
experimental data on the particle masses \cite{data}.

It is seen that the particle masses are solely determined by the value of 
$\alpha ^{'}.$ Although this parameter is known to coincide for both light 
mesons and baryons, it is also known to have a weak flavor dependence for 
light mesons, as remarked above. Since here we are not concerned with 
accuracies of better than 1\% (i.e., at the level of electromagnetic 
corrections), it would be enough to neglect the flavor dependence of 
$\alpha ^{'}$ and take 
\beq
\alpha ^{'}=0.85\;{\rm GeV}^{-2},
\eeq
which is the average of its values for $n\bar{n},$ $s\bar{n}$ and $s\bar{s}$
(see above). 

Let us start with (13). The use of $m_\pi =(m_\pi ^0+m_\pi ^{\pm })/2=137.3$
MeV \cite{data} in this formula leads, via (19), to
\beq
m_K=489.3\;{\rm MeV,}\;\;\;{\rm vs.}\;\;m_K=495.7\pm 2.0\;{\rm MeV}\;[7].
\eeq

Similar comparison of the hadron masses predicted by (15)-(18) with
data is presented in Table I.

\begin{center}
\begin{tabular}{|c|c|c|} \hline
 State & Mass from (15)-(18), MeV & Mass from ref. [7], MeV   \\ 
\hline
     $\rho $    &  767.0 & $768.5\pm 0.6$   \\ \hline
    $K^\ast $   &  899.3 & $893.9\pm 2.3$   \\ \hline
     $\phi $    & 1014.6 &     1019.4       \\ \hline
       $N$      &  939.3 &      938.9       \\ \hline
  $\Sigma ^{'}$ & 1150.5 &   $1155\pm 2$    \\ \hline
      $\Xi $    & 1328.4 &   $1318\pm 3$    \\ \hline
    $\Delta $   & 1212.7 &   $1232\pm 2$    \\ \hline
$\Sigma ^\ast $ & 1382.7 &   $1385\pm 2$    \\ \hline
  $\Xi ^\ast $  & 1533.9 & $1533.5\pm 1.5$  \\ \hline
    $\Omega $   & 1671.6 &     1672.4       \\ \hline
\end{tabular}
\end{center}
{\bf Table I.} Comparison of the particle masses predicted by the formulas 
(15)-(18) with experimental data from ref. \cite{data}.
 \\

One sees excellent agreement with experiment for all states, except for 
$\Delta .$ We however note that this state has largest width among the ground 
state baryons $(\sim 120$ MeV; for comparison, the $\Sigma ^\ast $ has largest
width of $\sim 9.5$ MeV among the remaining ground state baryons), and 
therefore its mass is poorly known. Indeed, the pole position of $\Delta ,$ as
indicated in \cite{data}, is $1210\pm 1$ MeV, and hence the prediction of Eq.
(18) for the $\Delta $ mass is in excellent agreement with the pole position
of $\Delta .$

One can easily obtain expressions similar to (15)-(18) for other 
hadronic multiplets, assuming that they populate liniar trajectories; 
e.g.,\footnote{It is interesting to note that, although the numerical values 
of $m^2_{a_2}=m^2_{\Xi }=3/(2\alpha ^{'}),$ as calculated from our formulas, 
do not coincide with data for $\alpha ^{'}=0.85$ GeV$^{-2},$ they do coincide 
with each other: $m_{a_2}=m_\Xi =1318$ MeV.} $m^2_{a_2}=m^2_{\rho }+1/\alpha 
^{'}=3/(2\alpha ^{'})=1328.4$ MeV, vs. $1318\pm 1$ MeV \cite{data}. 

\section{Concluding remarks}
We have shown that collinear Regge trajectories for mesons and baryons, and
the cubic mass spectrum associated with them, determine expressions of the 
hadron masses in terms of the universal Regge slope $\alpha ^{'}$ alone, which
are in excellent agreement with experiment for $\alpha ^{'}=0.85$ GeV$^{-2}.$ 

These expressions are consistent with a universal scaling behavior for all
hadron masses which has been widely discussed in the recent literature 
\cite{linNPA,scaling}. Indeed, as we show in a separate publication 
\cite{sep}, Regge phenomenology is consistent with the only form of such a 
scaling, $M^\ast /M=(\alpha ^{'}/\alpha ^{'\ast })^{1/2}$ (asterisk indicates 
a temperature- and/or density-dependent quantity), in view of which a relation
$M^2\propto 1/\alpha ^{'}$ is clearly uderstood. 

We note that the technique discussed in this paper can be easily deneralized
to glueballs \cite{glue}, and reproduces the scalar and tensor glueball masses
in excellent agreement with lattice QCD simulations. Further generalization of
this technique to multi-quark states (e.g., diquonia and pentaquarks) is of 
great interest, and will be undertaken elsewhere.


\bigskip
\bigskip
\begin{thebibliography}{9}
\bibitem{BL} S.Z. Belenky and L.D. Landau, Sov. Phys. Uspekhi {\bf 56} (1955) 
309; Suppl. Nuovo Cim. {\bf 3} (1956) 15
\bibitem{phen} E.V. Shuryak, Sov. J. Nucl. Phys. {\bf 16} (1973) 220
\bibitem{linNPA} L. Burakovsky and L.P. Horwitz, Nucl. Phys. A {\bf 614} 
(1997) 373
\bibitem{linmes}  L. Burakovsky and L.P. Horwitz, Found. Phys. Lett. {\bf 9} 
(1996) 561 
\bibitem{linbar}  L. Burakovsky and L.P. Horwitz, Found. Phys. Lett. {\bf 10} 
(1997) 61 
\bibitem{BHNPA} L. Burakovsky and L.P. Horwitz, Nucl. Phys. A {\bf 609}
(1996) 585
\bibitem{data} Particle Data Group (R.M. Barnett {\it et al.}), Phys. Rev. D
{\bf 54} (1996) 1
\bibitem{scaling} G.E. Brown and M. Rho, Phys. Rev. Lett. {\bf 99} (1991) 2720
\\ G.E. Brown, H.A. Bethe and P.M. Pizzochero, Phys. Lett. B {\bf 263} (1991) 
337 \\ K. Kusaka and W. Weise, Phys. Lett. B {\bf 288} (1992) 6 \\
G.E. Brown, A.D. Jackson, H.A. Bethe and P.M. Pizzochero, Nucl. Phys. A 
{\bf 560} (1993) 1035
\bibitem{sep} L. Burakovsky, Hadron Mass Scaling in Regge Phenomenology, LANL
preprint LA-UR-98-1843 [hep-ph/9805220]
\bibitem{glue} L. Burakovsky, Glueball Spectroscopy in Regge Phenomenology, 
LANL preprint LA-UR-98-1695 [hep-ph/9804465]
\end{thebibliography}
\end{document}